\documentclass[twocolumn, aps, showpacs,preprintnumbers,amsmath,amssymb]{revtex4}

\usepackage{graphicx}
\usepackage{dcolumn}
\usepackage{bm}
\usepackage[colorlinks,citecolor=blue,linkcolor=red]{hyperref}
\usepackage{color}
\usepackage{times}

\begin{document}


\title{Nonequilibrium Energy Transfer at Nanoscale: A Unified Theory from Weak to Strong Coupling}

\author{Chen Wang$^{1,2}$}
\author{Jie Ren$^{1,3,}$}\email{jieustc@gmail.com}
\author{Jianshu Cao$^{1,2,}$}\email{jianshu@mit.edu}

\address{
$^{1}$Department of Chemistry, Massachusetts Institute of Technology, 77 Massachusetts Avenue, Cambridge, MA 02139, USA\\
$^{2}$Singapore-MIT Alliance for Research and Technology, 1 CREATE Way, Singapore 138602, Singapore\\
$^{3}$Theoretical Division, Los Alamos National Laboratory, Los Alamos, New Mexico 87545, USA}

\date{\today}


\begin{abstract}
We investigate the microscopic mechanism of quantum energy transfer in the nonequilibrium spin-boson model.
By developing a nonequilibrium polaron-transformed Redfield equation based on fluctuation decoupling,
we dissect the energy transfer into multi-boson associated processes with even or odd parity.
Based on this, we analytically evaluate the energy flux, which smoothly bridges the transfer dynamics from the weak spin-boson coupling regime to the strong-coupling one.
Our analysis explains previous limiting predictions and provides a unified interpretation of several observations, including coherence-enhanced heat flux and absence of negative differential thermal conductance in the nonequilibrium spin-boson model. The results may find wide applications for the energy and information control in nanodevices.
\end{abstract}

\pacs{44.90.+c, 05.60.Gg, 44.10.+i, 63.22.-m}


\maketitle

\section{Introduction}

\begin{figure}[tbp]
\begin{center}
\vspace{-10mm}
\includegraphics[width=0.45\textwidth]{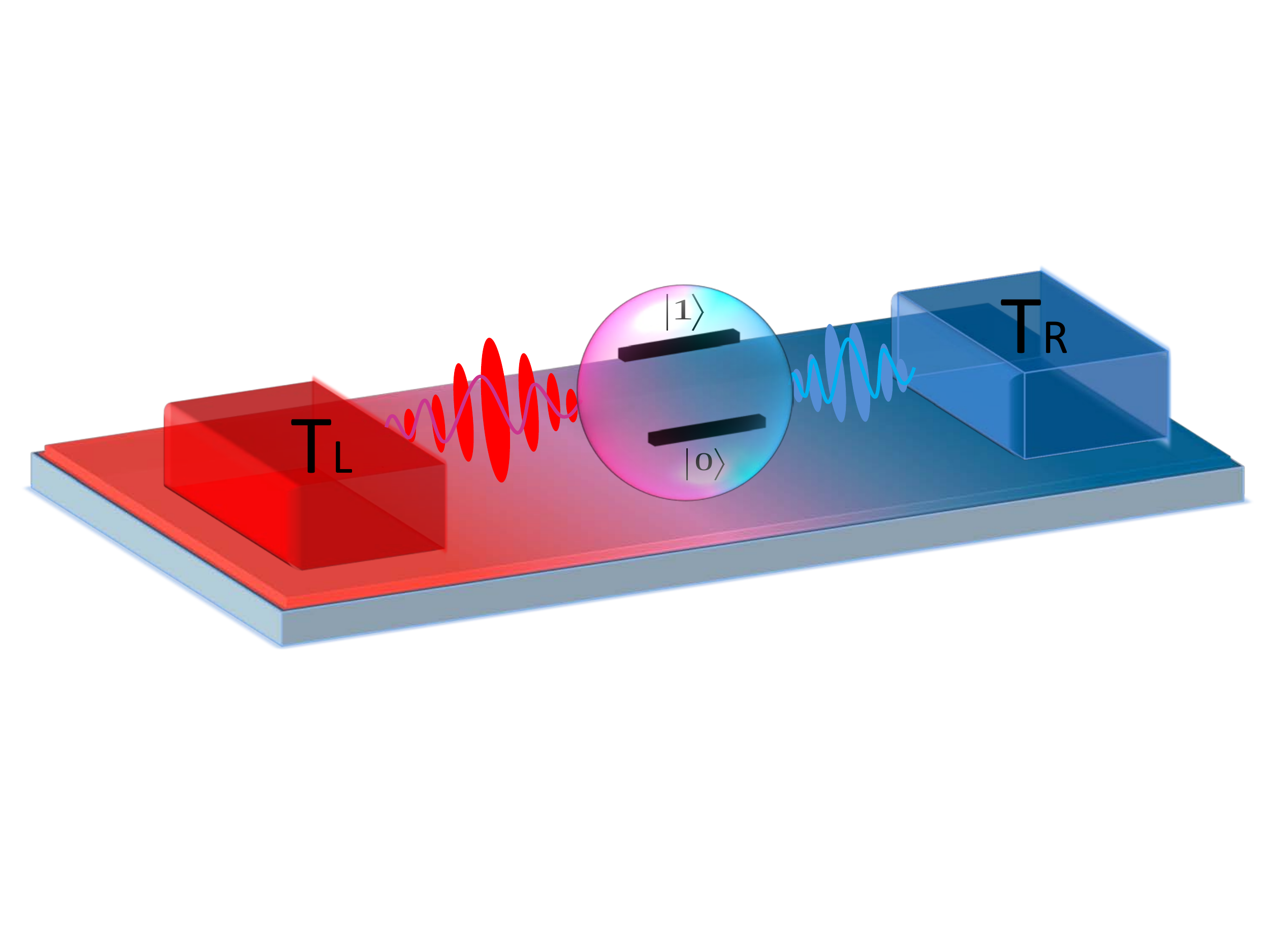}
\vspace{-10mm}
\end{center}
\caption{ Schematic illustration of the nonequilibrium spin-boson model composed by central two-level nanodevice connecting to two separate bosonic baths with temperature $T_L$ and $T_R$ respectively.}~\label{fig:fig01}
\label{fig1}
\end{figure}

Energy dissipation has become a severe bottleneck to the sustainability of any modern economy~\cite{Chu}.
To address this issue, efficient energy transfer and the corresponding smart control and detection at nanoscale have created unprecedented opportunities and challenges~\cite{nitzana1,mratner1,nbli1}. Therefore, understanding and controlling energy transfer in low-dimensional systems is of significant importance
not only in fundamental researches but also in practical applications~\cite{nitzana2,aishzaki1,jcao1,jianlanwu2}.
The simplest paradigm of nanoscale energy transfer is the spin-boson model,
regularly represented by a two-level system (TLS) interacting with a single bosonic bath, which has the equilibrium state after long time evolution.
It has been extensively investigated in quantum optics~\cite{moscully1}, quantum dissipation~\cite{aleggett1,uweiss1}, quantum phase transition~\cite{rbulla1,yyzhang1},
anomalous statistics~\cite{juzarthingna1,cklee1}, etc.
While for energy transport far from equilibrium, at least two baths should be included with thermodynamic bias (e.g. temperature bias), as shown in Fig. \ref{fig1}.
It results in the nonequilibrium spin-boson model (NESB), given by
\begin{equation}~\label{h1}
\hat{H}_0=\frac{\epsilon_0}{2}\hat{\sigma}_z+\frac{\Delta}{2}\hat{\sigma}_x+\hat H_B+\sum_{k;v=L,R}\hat{\sigma}_z(\lambda_{k,v}\hat{b}^{\dag}_{k,v}+\lambda^{*}_{k,v}\hat{b}_{k,v}),
\end{equation}
where the TLS is represented by spin matrices
$\hat{\sigma}_z=|1{\rangle}{\langle}1|-|0{\rangle}{\langle}0|$ and $\hat{\sigma}_x=|1{\rangle}{\langle}0|+|0{\rangle}{\langle}1|$,
with $\epsilon_0$ the energy spacing and $\Delta$ the tunneling strength  between the TLS.
$\hat H_B:=\sum_{k;v=L,R}\omega_{k,v}\hat{b}^{\dag}_{k,v}\hat{b}_{k,v}$ denotes the bosonic baths with $\hat{b}^{\dag}_{k,v} (\hat{b}_{k,v})$ creating (annihilating) one boson with energy $\omega_{k,v}$ and momentum $k$
in the $v$th bath. The last term describes the spin-boson interaction with $\lambda_{k,v}$ the coupling strength.
In the long time limit, the system reaches the nonequilibrium steady state (with stable energy flow), rather than the equilibrium state as described in the single bath spin-boson model case.

From NESB, the TLS can manifest itself as excitons, impurity magnets, anharmonic molecules, cold atoms, low-energy band structures and spins. And bosonic baths can register as electromagnetic environments, lattice vibrations, Luttinger liquid, magnons, and so on.
Hence, the NESB has already found widespread applications in fertile frontiers.
Particularly, in molecular electronics~\cite{mgalperin1,jcc1}  NESB describes electronic transport through a molecular junction.
In the donor-bridge-acceptor complexes, even if the bridged structure becomes long, such as proteins, the complexes can still reduce to the effective TLS description by
considering weak interactions of the bridge with the donor and acceptor~\cite{slarsson1,ssskourtis1}, which clearly illustrates that NESB can also describe large interesting systems.
In phononics~\cite{nbli1}, NESB describes the phononic energy transfer in anharmonic molecular junctions~\cite{dsegal1,dsegal3,dsegal2,dsegal4,renjie1,dsegal8}, and can be regarded as a special realization of the famous Caldeira-Leggett model~\cite{Caldeira}.
In many-body physics, NESB describes the novel Kondo physics and nonequilibrium phase transitions~\cite{LZhu,ksaito1}.
In spin caloritronics, NESB describes the nontrivial spin Seebeck effects that pave the way for thermal-driven spin diode and transistor~\cite{SSE}.
In quantum biology, NESB models the exciton transfer embedded in the photosynthetic complexes~\cite{aishzaki1,jianlanwu2,jianlanwu1,jmoix1,sfh1}.
Also, NESB describes electromagnetic transport through superconducting circuits~\cite{Niskanen} and photonic waveguides with a local impurity~\cite{LeHur}.
Moreover, this generic model can be extended to one dimensional spin chains at ultra-low temperatures~\cite{mvogl1}.

However, in spite of the wide range of applications of NESB, little is known about its underlying physics and the fundamental transfer mechanism.
To explore energy transfer in NESB, many approaches have been proposed,
but each approach works with limitations and is unable to see the wood for the trees. Typically, in the weak spin-boson coupling limit, the Redfield master equation with second order perturbation applies and gives the resonant energy transfer and the additive contributions of separate baths~\cite{dsegal1,renjie1}.
While in the strong spin-boson coupling limit,
the nonequilibrium version of the noninteracting-blip approximation (NIBA) equation applies and provides the off-resonant steady energy transfer and the non-additive picture~\cite{dsegal2,dsegal6,dsegal7,tianchen1},
which is usually based on the second order Born approximation in the polaron framework.
Note the traditional NIBA of a single bath spin-boson model is consistent with the Redfield scheme in the weak-coupling regime~\cite{aleggett1}, which is distinct from the NIBA results of the NESB that only apply in the strong-coupling regime~\cite{tianchen1}.
Moreover, for the negative differential thermal conductance (NDTC), the nonequilibrium NIBA scheme claims its appearance in the strong coupling for NESB, whereas the Redfield scheme predicts its absence in the weak coupling~\cite{dsegal2}. Similar limitations between these two schemes also occur in the high order flux-fluctuations as well as in the geometric-phase-induced energy transfer~\cite{tianchen1}.

Consequently, questions are raised: Which scheme provides the correct picture to describe the energy transfer in NESB? Or both are correct but are just two different manifestations of the same physics under different conditions? If so, what happens at the intermediate coupling regime?
To solve the long-standing challenge and answer these important questions, we propose a unified theory.
Although some numerical simulations have recently been carried out~\cite{ksaito1,kavelizhanin1,kavelizhanin2}, attempting to exactly calculate the energy transfer in NESB,
they all have their own limitations or require expensive computations.
Moreover, numerical approaches may not provide clear physical insights to the underlying energy transfer mechanism.
Therefore, it is indeed crucial to develop a unified analytical theory in order to fully resolve the microscopic mechanism and uncover new physics of the energy transfer in NESB.

In this work, we firstly present a nonequilibrium polaron-transformed Redfield equation (NE-PTRE) in Sec. II, which based on the fluctuation-decoupling method perturbs the spin-boson interaction
in the polaron framework. This approach is capable of bridging the energy transfer pictures of NESB from weak to strong coupling regimes.
In Sec. III, we clearly unravel the energy transfer in NESB as multi-boson processes, which are classified by the odd-even parity, with the sequential- and co-tunneling behaviors as two lowest order contributions.
To further exemplify the power of our unified theory in Sec. IV, we derive the analytical expression of energy flux that dissects the transfer processes systematically, and show that this unified flux expression reduces to the NIBA at strong coupling limit and to the Redfield one at the weak coupling limit, respectively.
In Sec. V, we investigate NDTC and identify its absence over wide range of the temperature bias, even in the intermediate and strong coupling regimes,
which corrects the previous observation of NDTC under the NIBA in the classical limit~\cite{dsegal2}.
Finally, the conclusion is given in Sec. VI.

\section{Model and Method}

We first apply the canonical transformation to obtain the transformed Hamiltonian, which re-organizes the system-bath interaction beyond the billinear form.
Then, a non-equilibrium polaron-transformed based master equation is derived, which decouples the system-bath interaction non-perturbatively.

\subsection{Transformed Hamiltonian}

Based on the canonical transformation $\hat{U}=e^{i\hat{\sigma}_z\hat{B}/2}$ to the NESB Hamiltonian at Eq.~({\ref{h1}}),
a new transformed Hamiltonian is obtained as
\begin{eqnarray}
\hat{H}=\hat{U}^{\dag}\hat{H}_0\hat{U}=\frac{\epsilon_0}{2}\hat{\sigma}_z+\hat{H}_B+\hat{V}_{SB},
\end{eqnarray}
where the system-bath interaction is given by
\begin{eqnarray}~\label{sb1}
\hat V_{SB}=\frac{\Delta}{2}(\hat{\sigma}_x\cos\hat{B}+\hat{\sigma}_y\sin\hat{B})
\end{eqnarray}
with $\hat{B}=2i\sum_{k;v=L,R}(\frac{\lambda_{k,v}}{\omega_{k,v}}\hat{b}^{\dag}_{k,v}-\frac{\lambda^*_{k,v}}{\omega_{k,v}}\hat{b}_{k,v})$ the collective momentum operator of bosonic baths.
Traditionally, many methods, including the NIBA~\cite{dsegal6,dsegal7}, directly treat the interaction $\hat V_{SB}$ as a perturbation. However, we note that generally $\hat V_{SB}$ can not behave as a perturbation due to the non-negligible expectation $\langle \hat V_{SB} \rangle$,  except for weak inter-site tunneling or strong system-bath coupling.  Nevertheless, the fluctuation around the expectation $\hat V_{SB}-\langle \hat V_{SB} \rangle$ can be safely treated by the second order perturbation, regardless of the tunneling and coupling strength.  This is the key idea of the fluctuation decoupling scheme. Notice the thermal average $\langle \sin\hat B \rangle=0$, so that $\langle\hat V_{SB} \rangle=\eta\Delta\hat\sigma_x/2$, with the renormalization factor
\begin{eqnarray}~\label{eta1}
\eta=\langle \cos\hat B \rangle=\exp{(-\sum_{v}\int^{\infty}_0d{\omega}\frac{J_v(\omega)}{\pi\omega^2}[n_v(\omega)+1/2])},
\end{eqnarray}
where $J_v(\omega)=4\pi\sum_k|\lambda_{k,v}|^2\delta(\omega-\omega_k)$ is the $v$th bath spectrum and $n_v(\omega)=1/[\exp(\beta_v\omega_v)-1]$ denotes the corresponding Bose-Einstein distribution with $\beta_v=1/k_bT_v$ the inverse of the temperature. In energy transfer studies, the spectrum can be usually considered as $J_v(\omega)=\pi\alpha_v{\omega^{s}}{\omega_{c,v}^{1-s}}e^{-\omega/\omega_{c,v}}$ with $\alpha_v$ ($\sim|\lambda_{k,v}|^2$) the coupling strength and $\omega_{c,v}$ the cutoff frequency.
Without loss of generality, we choose the typical super-Ohmic spectrum $s=3$ for consideration in this paper~\cite{uweiss1}.
Then, the renormalization factor at Eq.~(\ref{eta1}) is specified as
$\eta=\exp\{\sum_{v=L,R}-\frac{\alpha_v}{2}[-1+\frac{2}{(\beta_v\omega_{c,v})^2}\psi_1(\frac{1}{\beta_v\omega_{c,v}})]\}$,
where the special function $\psi_1(x)=\sum^{\infty}_{n=0}\frac{1}{(n+x)^2}$ is the trigamma function.

We carry out the fluctuation-decoupling by subtracting the interaction by its expectation and compensating to the system Hamiltonian. As such, the transformed Hamiltonian is re-grouped as $\hat{H}=\hat H_S +\hat{H}_B+\hat V_{SB}$,
with the system Hamiltonian
\begin{eqnarray}~\label{h2}
\hat H_S=\frac{\epsilon_0}{2}\hat{\sigma}_z+\eta\frac{\Delta}{2}\hat{\sigma}_x,
\end{eqnarray}
and the new system-bath interaction
\begin{eqnarray}
\hat V_{SB}=\frac{\Delta}{2}[\hat{\sigma}_x(\cos\hat{B}-\eta)+\hat{\sigma}_y\sin\hat{B}].
\end{eqnarray}
Clearly, the renormalization factor $\eta$ approaches to $1$ at weak coupling $\alpha_v$ but vanishes to $0$ at the strong coupling limit. Therefore, by means of the fluctuation-decoupling, the new system-bath interaction can be reliably perturbed regardless of the coupling strength.

It should be noted that if we select the bosonic baths as the Ohmic case $s=1$, the renormalization factor expressed at Eq.~(\ref{eta1}) will always approach to zero regardless of the system-bath coupling strength,
and the expectation of the system-bath interaction at Eq.~(\ref{sb1}) ${\langle}\hat{V}_{SB}{\rangle}=0$.
As such, the NE-PTRE based on the fluctuation-decoupling scheme,
will be equivalent to the nonequilibrium NIBA~\cite{dsegal6,dsegal7,tianchen1}.

\subsection{Fluctuation-Decoupling based master equation}
Fluctuation-decoupling is the key step, based on which we are able to apply various perturbative methods to proceed.
Here, we adopt the nonequilibrium polaron-transformed Redfield equation and obtain (see Appendix \ref{sectionA})
\begin{equation}~\label{qme1}
\frac{{\partial}\hat{\rho}}{{\partial}t}=-i[\hat{H}_S,\hat{\rho}]
+\sum_{l=e,o}\sum_{\omega,\omega_1=0,{\pm}\Lambda}\Gamma_l(\omega)[\hat{P}_l(\omega)\hat{\rho},\hat{P}_l(\omega_1)]+H.c.  ,
\end{equation}
where $\hat{\rho}$ is the reduced density matrix for the TLS in the polaron framework, $\Lambda=\sqrt{\epsilon_0^2+\eta^2 \Delta^2}$ is the energy gap of the renormalized TLS in its eigenspace, and
$\hat{P}_{e(o)}(\omega)$ is the measuring projector in the eigen-basis obtained from the evolution of spin matrices
$\hat{\sigma}_{x(y)}(-\tau)=\sum_{\omega=0,{\pm}\Lambda}\hat{P}_{e(o)}(\omega)e^{i\omega\tau}$. The subscript $e (o)$ denotes the even (odd) parity of transfer dynamics. $\Gamma_{l}(\omega)$ with $l=e, o$ has the meaning of transition rate that we will discuss later in detail.

We note that R. J. Silbey and his colleagues have applied a similar polaron-based master equation to investigate the coherent exciton dynamics~\cite{rjsilbey1,rjsilbey11}. However, the system they studied interacts with merely a single bosonic bath, i.e. the standard equilibrium spin-boson model. Only with multiple baths, we are able to determine the underlying picture of energy transport through NESB to be resonant or off-resonant, additive or non-additive.
More importantly, this NE-PTRE provides us analytical solutions, which clearly demonstrate the energy transfer as multi-boson processes that are classified by odd-even parity.

\begin{figure}[t]
\begin{center}
\vspace{-10mm}
\includegraphics[width=0.45\textwidth]{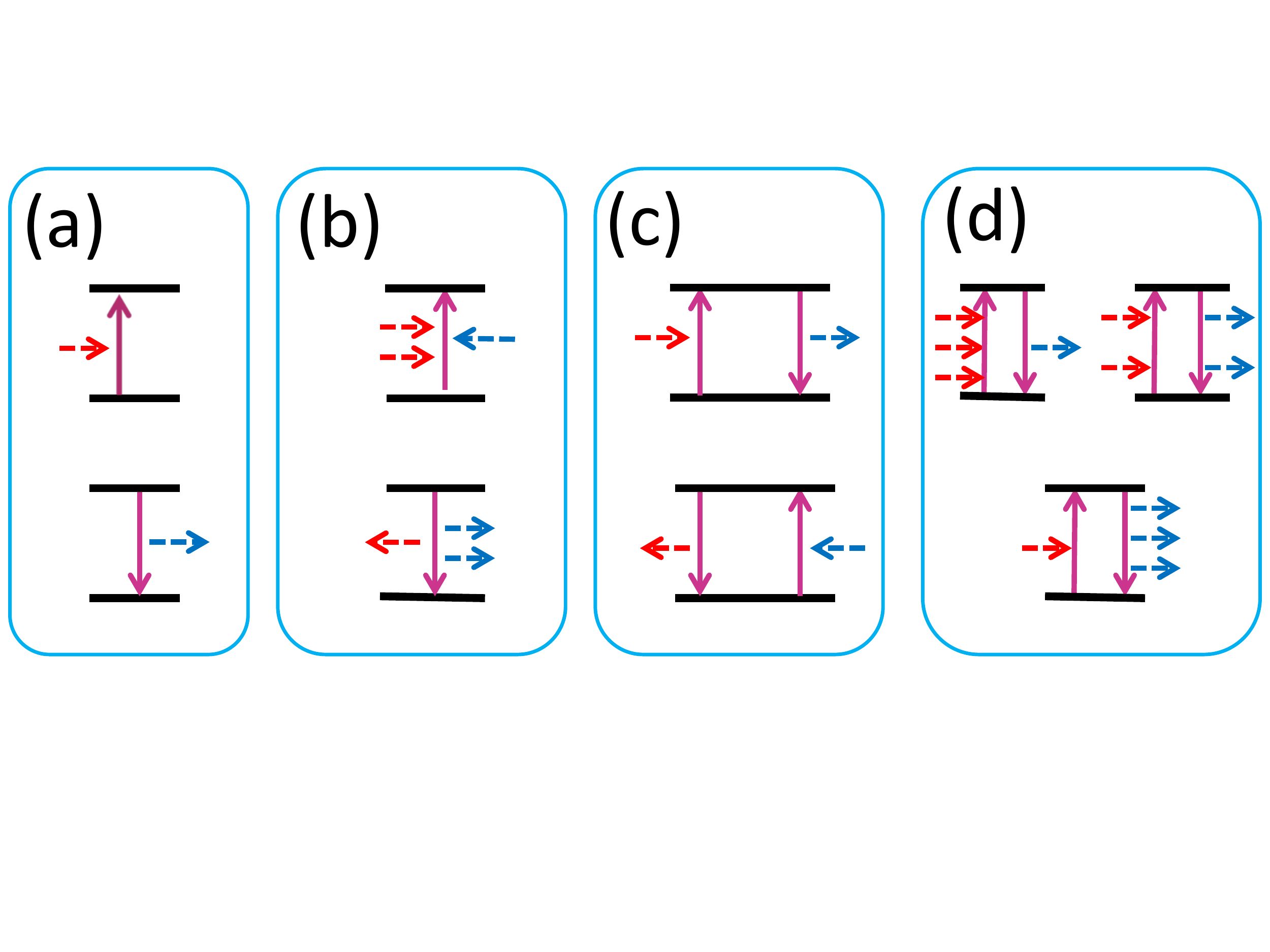}
\vspace{-20mm}
\end{center}
\caption{ Representative processes in multi-boson assisted energy transfer:
(a) single boson involved sequential process;
(b) three-boson involved ``tri-tunneling" process;
(c) two-boson ``cotunneling" process;
(d) four-boson involved collective process.}~\label{fig2}
\end{figure}


\section{Parity classified transfer processes}
As the crucial observation, the transition rates are expressed as  $\Gamma_{o,e}(\omega)=(\frac{\eta\Delta}{2})^2\int^{\infty}_0d{\tau}e^{i\omega\tau}\gamma_{o,e}(\tau)$,
where the correlation functions are specified by (see Appendix \ref{sectionB})
\begin{eqnarray}~\label{q1}
\gamma_o(\tau)&=&{\sinh}[Q(\tau)]=\sum^{\infty}_{n=0}\frac{Q(\tau)^{2n+1}}{(2n+1)!},    \nonumber\\
\gamma_e(\tau)&=&{\cosh}[Q(\tau)]-1=\sum^{\infty}_{n=1}\frac{Q(\tau)^{2n}}{(2n)!}.
\end{eqnarray}
The boson propagator $Q(\tau)=\sum_{v=L,R}Q_v(\tau)$ with
$Q_v(\tau)=\int^{\infty}_0d\omega \frac{J_v(\omega)}{\pi\omega^2}[n_v(\omega)e^{i\omega \tau}+(1+n_v(\omega))e^{-i\omega \tau}]$ describes the bosonic absorptions and emissions that constitute the energy transfer.
Clearly, the multi-boson processes are classified by the odd and even propagators,  with each order fully captured by
the corresponding Taylor expansion systematically.

Specifically, $\gamma_o(\tau)$ describes the processes involving odd boson numbers.
The lowest order contribution is the sequential-tunneling [see Fig. \ref{fig2}({a})], expressed by
$\Gamma^{(1)}_o(\omega)=\frac{(\eta{\Delta})^2}{8}(Q_L(\omega)+Q_R(\omega))$~\cite{renjie1,dsegal1,dsegal2},
with $Q_v(\omega)=\int^{\infty}_{-\infty}d{\tau}e^{i\omega\tau}Q_v(\tau)$ and $\omega={\pm}\sqrt{\epsilon_0^2+4\eta^2\Delta^2}$.
This means that the relaxation and excitation of the TLS is influenced by the $L$ and $R$ baths separately, i.e., additively.
Further, the higher order, called as "tri-tunneling" [Fig. \ref{fig2}({b})], is exhibited as
$\Gamma^{(2)}_o(\omega)=(\frac{\eta{\Delta}}{4\pi})^2\int{\int}d{\omega_1}d{\omega_2}\sum_v
Q_v(\omega_1)Q_v(\omega_2)Q_{\overline{v}}(\omega-\omega_1-\omega_2)$, with
$\overline{v}=L(R)$ for $v=R(L)$, where the baths act non-additively and off-resonantly.
This highly non-trivial term explicitly demonstrates the collective transfer process with different contributions from two baths.

Correspondingly, $\gamma_e(\tau)$ describes processes of even boson number participating in the energy transfer processes.
The lowest order includes the co-tunneling effect~\cite{truokola1} [see Fig. \ref{fig2}({c})].
It contributes to the transition rate as
$\Gamma^{(1)}_e(0)=\frac{(\eta{\Delta})^2}{8\pi}\int^{\infty}_0d{\omega_1}Q_L(\omega_1)Q_R(-\omega_1)$.
This implies that when the left bath releases energy $\omega_1$, the right bath absorbs the same quanta simultaneously,
leaving the TLS unchanged.
Clearly, two baths are involved non-additively.
The corresponding higher order term can also be obtained systematically [see Fig. \ref{fig2}({d})].
As a result, we can dissect the contribution of each order of boson excitations to the energy transfer
based on the expansions, and the underlying multi-boson transfer mechanism can be systematically exploited.

\begin{figure}[t]
\begin{center}
\vspace{-3.6cm}
\includegraphics[width=0.55\textwidth]{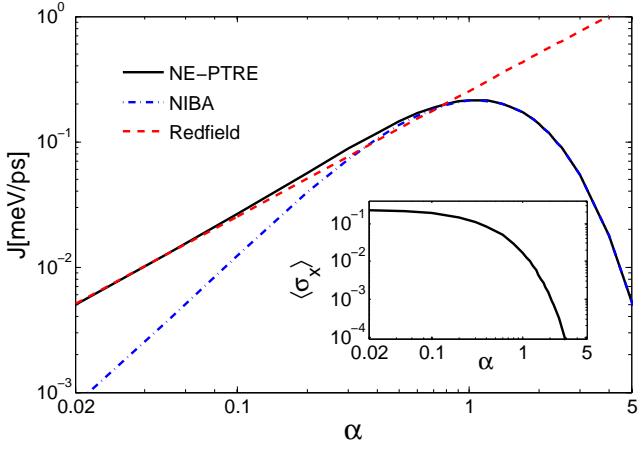}
\vspace{-3.5cm}
\end{center}
\caption{ The energy flux and quantum coherence represented by $\langle\sigma_x\rangle$,
as functions of the coupling strength.
The solid black line is from the NE-PTRE, which unifies the Redfield result at the weak coupling (the red dashed line) and the NIBA result at the strong coupling (the dot-dashed blue line). The deviation of the unified energy flux from the NIBA result at small $\alpha$ is characterized by the quantum coherence $\langle\sigma_x\rangle$ (inset).
Parameters are given by $\epsilon_0=0$, $\Delta=5.22$~meV, $\omega_c=26.1$~meV, $T_L=150$~K, and $T_R=90$~K.}~\label{fig3}
\end{figure}

\section{Unified energy flux from weak to strong couplings}
To exploit the dynamical processes corresponding to the correlation functions in Eq.~({\ref{q1}}),
we introduce the rate $\phi_{e(o)}(\omega)=\int^{\infty}_{-\infty}d{\tau}e^{i\omega\tau}\gamma_{e(o)}(\tau)$ in the frequency domain.
As such, when rewriting $\phi_{e(o)}(\omega)=\frac{1}{2\pi}\int^{\infty}_{-\infty}d{\omega^{\prime}}C_{e(o)}(\omega,\omega^{\prime})$, we are able to extract the corresponding kernel functions (see Appendix \ref{sectionB})
\begin{eqnarray}~\label{cxy1}
C_e(\omega,\omega^{\prime})&=&\frac{1}{2}\sum_{\sigma={\pm}}C^{\sigma}_{L}(\omega-\omega^{\prime})C^{\sigma}_R(\omega^{\prime})-\delta(\omega^{\prime}),\nonumber\\
C_o(\omega,\omega^{\prime})&=&\frac{1}{2}\sum_{\sigma=\pm}{\sigma}C^{\sigma}_L(\omega-\omega^{\prime})C^{\sigma}_R(\omega^{\prime}).
\end{eqnarray}
where $C^{\pm}_v(\omega^{\prime})={\int^{\infty}_{-\infty}}d{\tau}e^{i\omega^{\prime}\tau{\pm}Q_v(\tau)}$ describes the rate density of the $v$th bath absorbing (emitting) energy $\omega$~$(-\omega)$, obeying the detailed balance relation as
$C^{\pm}_v(\omega^{\prime})/C^{\pm}_v(-\omega^{\prime})=e^{\beta_v\omega^{\prime}}$.

These kernel functions provide the other way of understanding the odd-even parity assisted energy tunneling processes that incorporate two baths non-additively.
Physically, $C_{e(o)}(\omega,\omega^{\prime})$ describes that when the TLS releases energy $\omega$
by relaxing from the excited state to the ground one,
the right bath absorbs energy $\omega^{\prime}$ and the left one obtains the left $\omega-\omega^{\prime}$ if $\omega>\omega^{\prime}$ or supply the compensation if $\omega<\omega^{\prime}$.
And $C_{e(o)}(-\omega,\omega^{\prime})$ describes similar dynamical processes for the TLS jumping from the ground state to the exciting one.
While $\phi_{e(o)}(\omega)$ is the summation behavior of these corresponding microscopic processes.

Next, combined with the counting field~\cite{espositom1}, the nonequilibrium Polaron-transformed Redfield master equation in the Liouville space is shown as
$\frac{d|\rho_{\chi}(t){\rangle}{\rangle}}{dt}=-\hat{\mathcal{L}}_{\chi}|\rho_{\chi}(t){\rangle}{\rangle}$, with the column vector form of the reduced system density matrix
$|\rho_{\chi}(t){\rangle}{\rangle}=[{\langle}1|\hat{\rho}_{\chi}(t)|1{\rangle},{\langle}0|\hat{\rho}_{\chi}(t)|0{\rangle},{\langle}1|\hat{\rho}_{\chi}(t)|0{\rangle},
{\langle}0|\hat{\rho}_{\chi}(t)|1{\rangle}]^{T}$, and $\hat{\mathcal{L}}_{\chi}$ the Liouville super-operator.
Hence, the expression of the steady energy flux is given by
$\mathcal{J}={\langle}{\langle}1|\frac{{\partial}\hat{\mathcal{L}}_{\chi}}{{\partial}i\chi}|_{\chi=0}|\Psi{\rangle}{\rangle}$ (see Appendix \ref{sectionC}),
with ${\langle}{\langle}1|=[1,1,0,0]$, and $|\Psi{\rangle}{\rangle}$ the traditional steady state of the system density matrix ($\chi=0$).

In many energy transfer studies, the resonant case ($\epsilon_0=0$) is of the prime interest.
The steady state can be obtained in the local basis as (see Appendix \ref{sectionD})
\begin{eqnarray}
P_{11}=P_{00}=1/2,~P_{01}=P_{10}=\frac{1}{2}\frac{\phi_o(-\Lambda)-\phi_o(\Lambda)}{\phi_o(-\Lambda)+\phi_o(\Lambda)}
\end{eqnarray}
with the energy gap $\Lambda=\eta\Delta$ and $P_{ij}={\langle}i|\hat{\rho}(\infty)|j{\rangle}$.
Finally, we obtain the energy flux as
\begin{eqnarray}~\label{j1}
\mathcal{J}&=&\frac{{\Lambda}^2}{8\pi}\int^{\infty}_{-\infty}d{\omega}{\omega}
\bigg[\frac{\phi_o(\Lambda)C_o(-\Lambda,\omega)+\phi_o(-\Lambda)C_o(\Lambda,\omega)}{\phi_o(\Lambda)+\phi_o(-\Lambda)}\nonumber\\
&&+C_e(0,\omega)\bigg].
\end{eqnarray}
It is interesting to find that in the odd number parity subspace, as the TLS relaxes energy $\Lambda$, the baths show collective contribution
$C_o(-\Lambda,\omega)$ to the flux with the weight $\phi_o(\Lambda)/\sum_{\sigma=\pm}\phi_o(\sigma\Lambda)$.
Similarly, when the TLS is excited by an energy $\Lambda$, $C_o(\Lambda,\omega)$ is contributed to the flux with the corresponding subspace weight as $\phi_o(-\Lambda)/\sum_{\sigma=\pm}\phi_o(\sigma\Lambda)$.
While for the even parity subspace, the TLS energy is unchanged, with the contribution $C_e(0,\omega)$ to the flux.
This unified energy flux expression clearly uncovers that two parity-classified sub-processes both contribute to the energy transfer, whereas
the Redfield approach merely includes the lowest odd order and the NIBA only considers the even order.


Analytically, in the weak coupling limit, one only needs to keep the leading order of the correlation function as $O(\alpha_v)$ so that the renormalization factor is simplified to $\eta\approx1$ and $\Lambda=\Delta$.
Hence, the kernel function with even parity $C_e(0,\omega)=0$ and the odd one becomes
$C_o({\pm}\Delta,\omega)=2\pi[\delta({\pm}\Delta-\omega)Q_R({\pm}\Delta)+\delta(\omega)Q_L({\pm}\Delta)]$.
The unified energy flux reduces to the resonant energy transfer
\begin{eqnarray}
\mathcal{J}_w=\frac{\Delta}{2}\frac{J_L(\Delta)J_R(\Delta)(n_L-n_R)}{J_L(\Delta)(1+2n_L)+J_R(\Delta)(1+2n_R)},
\end{eqnarray}
with $n_v=n_v(\Delta)$, which is consistent with previous results of Redfield approach~\cite{renjie1,dsegal1}.
While in the strong coupling limit, multiple bosons are excited from baths, and both the renormalization factor $\eta$ and the eigen-energy gap of the TLS $\Lambda$ become zero.
Hence, two subspace kernel functions at Eq.~({\ref{cxy1}}) show equal weight.
The energy flux can be finally expressed as
\begin{eqnarray}~\label{js1}
\mathcal{J}_s=\frac{({\Delta}/2)^2}{2\pi}\int^{\infty}_{-\infty}d{\omega}{\omega}C_{L}(-\omega)C_R(\omega),
\end{eqnarray}
with the probability density of the $v$th bath
\begin{eqnarray}~\label{proden1}
C_v(\omega)&=&\int^{\infty}_{-\infty}\!\!\!\!\!\exp{[i{\omega}t+Q_v(t)-{\int}d{\omega^{\prime}}
\frac{J_v(\omega^{\prime})}{\pi{\omega^{\prime}}^2}\coth\frac{\beta_v\omega^{\prime}}{2}]dt},\nonumber\\
\end{eqnarray}
which correctly recovers the nonequilibrium NIBA result~\cite{dsegal2,dsegal6,dsegal7}.

\begin{figure}[t]
\begin{center}
\vspace{-1.2cm}
\includegraphics[width=0.5\textwidth]{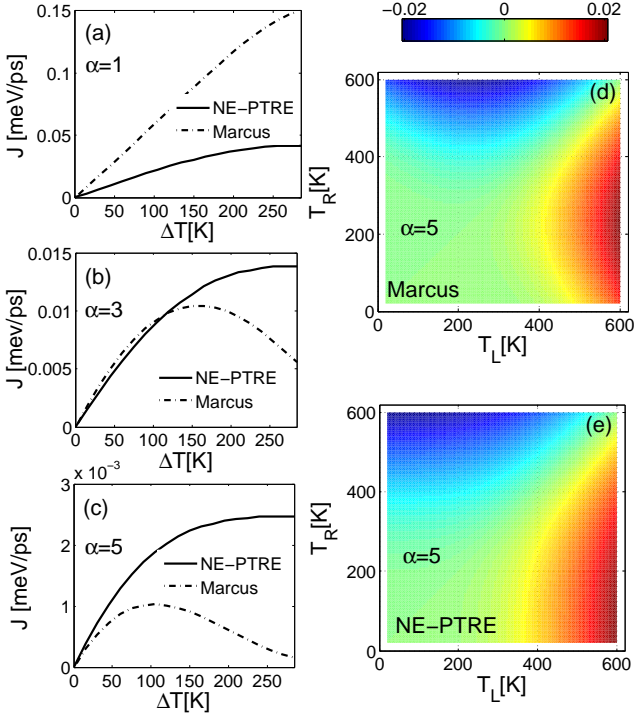}
\vspace{-0.5cm}
\end{center}
\caption{ Energy flux in the intermediate and strong system-bath coupling regimes by tuning the right bath temperature in (a-c),
and the birdeye view of the energy flux by varying the two bath temperatures in (d-e).
The parameters are given by $\Delta=1.0$~meV, $\epsilon_0=10$~meV, $\omega_c=26.1$~meV,
$T_L=300$~K, and $T_R=T_L-{\Delta}T$.}~\label{fig4}
\end{figure}

Next, we plot the energy flux of Eq.~({\ref{j1}}) in Fig.~\ref{fig3},
which first shows linear increase with the system-bath coupling at weak regime, consistent with the Redfield.
After reaching a maximum, the energy flux decreases monotonically in the strong coupling regime,
of which the profile coincide with the NIBA. The discrepancy of the NIBA and our NE-PTRE is due to the improper ignorance of quantum coherence $\langle\sigma_x\rangle$ of the TLS in NIBA (see also Eq.~(\ref{h2}), in which the term containing $\sigma_x$ is absent in the NIBA method).
This coherence term describes the effective tunneling within TLS so that it enhances the energy transfer compared to the NIBA that ignores it.

Therefore, we conclude that the unified energy flux expression of Eq.~({\ref{j1}}) provides a comprehensive interpretation
for energy transfer in NESB, because the fluctuation-decoupling not only describes the coherent system-bath coupling from weak to strong regimes, but also correctly captures the coherence within the TLS.

\section{Absence of negative differential thermal conductance}

NDTC, a typical feature in energy transport, has
been extensively studied in phononic devices~\cite{nbli1}.
In particular, NDTC has also been exploited in molecular junctions,
represented by the NESB.
By adopting nonequilibrium NIBA in the Marcus limit, i.e. high temperature baths, it was reported that NDTC is absent in the weak coupling but emerges in the strong coupling regime~\cite{dsegal2}.
However, what happens at the intermediate coupling regime is unclear. Moreover, it is questionable that whether NDTC is still presented in the comparatively low temperature regime.

Marcus theory was originally proposed to study semi-classical electron transfer rates in  the donor-acceptor species~\cite{uweiss1,ramarcus1}.
In previous works of energy transport~\cite{dsegal6,dsegal7,dsegal8,tianchen1}, the system dynamics with the Marcus limit is described by the rate equation based on the nonequilibrium NIBA, i.e. Eq.~(3) in Ref.~\cite{tianchen1}.
In the high temperature limit, it is known that $n_v(\omega){\approx}1/(\beta_v\omega)$ and the low frequency domain of bosonic baths dominates the evolution, which corresponds to the short-time expansion $1-\cos\omega\tau\approx\omega^2\tau^2/2$ and $\sin\omega\tau{\approx}\omega\tau$~\cite{dsegal1,dsegal2,tianchen1}. Thus the Gaussian decay of the the probability density is given by
$C_v(t)=\sqrt{\frac{\pi\beta_v}{\Gamma_v}}\exp[-\frac{\beta_v(\omega-\Gamma_v)^2}{4\Gamma_v}]$, which is as the same as derived from Eq.~(\ref{proden1}) even under biased condition.
The renormalized coupling strength is $\Gamma_v=\int^\infty_0d\omega\frac{J_v(\omega)}{\pi\omega}=4\sum_k|\lambda_{k,v}|^2/\omega_k$.
Hence, the energy flux can be obtained accordingly with the help of counting field~\cite{tianchen1}.
However, this limiting picture may be modified if the temperatures of bosonic baths become low,
when the quantum effect will be included to make the probability density non-Gaussian.

Therefore, we re-examine the NDTC by the NE-PTRE in Fig.~\ref{fig4}({a}-{c}).
In the intermediate coupling regime ($\alpha=1$), the energy flux increases monotonically by enlarging the temperature bias ($\Delta{T}=T_L-T_R$), both for the NE-PTRE and the NIBA of the classical (Marcus) limit.
As the coupling is strengthened further into the strong coupling regime (i.e. $\alpha=3$ and $\alpha=5$),
NDTC is apparent at the Marcus limit~\cite{dsegal2}.
However, no turnover signal is found based on the NE-PTRE.
In the strong coupling limit, $\eta\rightarrow0$ so that our method reduces to the NIBA, thus the discrepancy comes from the Marcus approximation.
It should be noted that from Figs. \ref{fig4}b and \ref{fig4}c, the qualitative deviation occurs at the large temperature bias.
This means that the temperature of the right bath is rather low, and quantum effect as such low temperatures may change behaviors of the correlation functions.
To further clarify the absence of NDTC at the deep strong coupling regime,
the birdeye view contours of energy flux are compared with and without Marcus limit [see Fig. \ref{fig4}({d}) and \ref{fig4}({e})].
It is shown that the turnover behavior appears within the Marcus framework, by tuning either $T_L$ or  $T_R$ (see Fig. 4({d})), whereas it never emerges with rigorous calculations [see Fig. \ref{fig4}({e})]. In fact, this result clearly demonstrates that the NDTC in the Marcus limit occurs at large temperature bias $\Delta T=T_L-T_R$ with either $T_R$ or $T_L$ at very low temperature, where the high temperature precondition of the Marcus framework may break down. Thus the NDTC observed in the NIBA scheme with Marcus assumption is merely an artifact.

Hence, we conclude that by tuning one bath temperature, NDTC is absent across a wide range of the temperature bias in the NESB, even in the strong system-bath coupling limit. Finally, we would like to note if we allow to change two temperatures simultaneously, NDTC can still occur in NESB. Also, NDTC is not exclusive to the strong coupling limit generally, but can even exist in the weak coupling regime if  the system is hybridized with fermion-spin-boson couplings~\cite{SSE, renjie3}.

\section{Conclusion}

By applying the nonequilibrium polaron-transformed Redfield equation based on fluctuation decoupling, we have unified the energy transfer mechanisms in the nonequilibrium spin-boson model from weak to strong coupling regimes. Specifically, we have characterized energy transfer as multi-boson processes that are classified by the odd-even parity. We have analytically obtained the energy flux expression in Eq.~({\ref{j1}}),
which explicitly unifies the analytic results from the weak-coupling Redfield scheme and the strong-coupling NIBA scheme.
Moreover, enhancement of the energy flux at the intermediate coupling regime has been identified, which results from the persistence of coherent tunneling within the TLS but is unexpectedly ignored in the nonequilibrium NIBA. Other relevant limiting problems of energy transfer have also been systematically resolved. Our analytic and numeric results provide a comprehensive interpretation of previous works, fully resolve the microscopic mechanism of energy transfer, and should have broad implications for smart control of energy and information in low-dimensional nanodevices.

\begin{acknowledgments}
This work was supported by the National Science Foundation (NSF) (grant no. CHE-1112825) and Defense Advanced Research Projects Agency (DARPA) (grant no. N99001-10-1-4063).
C. Wang has been supported by Singapore-MIT Alliance for Research and Technology (SMART).
The early work of J. Ren was supported by the National Nuclear Security Administration of the U.S. DOE at LANL under Contract No. DE-AC52-06NA25396, through the LDRD Program.
J. Cao has been partly supported by the Center for Excitonics,
an Energy frontier Research Center funded by the U.S. Department of Energy,
Office of Science, Office of Basic Energy Science.
\end{acknowledgments}

\appendix

\section{Nonequilibrium polaron-transformed Redfield equation and the auxiliary counting statistics}\label{sectionA}
Generally, the model of a quantum system interacting with two separate baths can be described by
\begin{equation}~\label{sh1}
\hat{H}=\hat{H}_s+\sum_v\hat{H}^v_b+\hat{V},
\end{equation}
where $\hat{H}_s$ denotes the system Hamiltonian, $\hat{H}^v_b$ models the $v$ bath,
and $\hat{V}=\sum_v\hat{V}_v$ with $\hat{V}_v$ being the interaction between the system and the $v$ bath.
If the strength of the system-bath coupling is weak compared to the intrinsic energy scale of the system, Born-Markov approximation is usually applied to derive the second-order master equation~\cite{uweiss1}:
\begin{equation}~\label{sme1}
\frac{{\partial}\hat{\rho}_s(t)}{{\partial}t}=\frac{1}{i}[\hat{H}_s,\hat{\rho}_s(t)]
-\int^{\infty}_0d{\tau}\textrm{Tr}_b\{[\hat{V},[\hat{V}(-\tau),\hat{\rho}_s(t){\otimes}\hat{\rho}_b]]\},
\end{equation}
where $\hat{\rho}_s$ is the reduced system density operator, $\hat{\rho}_b=e^{-\sum_v\beta_v\hat{H}^v_b}/Z$ is the canonical distribution of the baths, and $\textrm{Tr}_b\{\cdot\}$ traces off the baths freedom degree.
As the (temperature or voltage) bias is applied on two baths, the quantum system is driven from the equilibrium state to the nonequilibrium steady state, which spontaneously generates the energy/particle flux.

In the following, we will generalize the quantum master equation by including the auxiliary counting field $\chi$. When setting $\chi=0$, everything will reduce to the conventional quantum master equation and the contents in the main text.
The counting statistics~\cite{snazarovyv1,slevitov1,slevitov2}, as a mathematically rigorous method~\cite{svaradhan1}, is usually applied to measure the arbitrary order of the energy current fluctuation~\cite{espositom1,scampisim1,dsegal6,dsegal7}, of which the lowest order gives the steady state flux.
If we count the number of particles into the $v$ bath, the Hamiltonian at Eq.~(\ref{sh1}) under the counting field is given by
$\hat{H}_{\chi}=e^{i{\chi}\hat{H}^v_b/2}\hat{H}e^{-i{\chi}\hat{H}^v_b/2}=\hat{H}_s+\sum_v\hat{H}^v_b+\hat{V}_{\chi}$.
Hence, the quantum master equation under the counting field $\chi$ is shown as~\cite{espositom1,swangc1,syuget1}
\begin{eqnarray}~\label{sme2}
\frac{{\partial}\hat{\rho}_{\chi}(t)}{{\partial}t}&=&\frac{1}{i}[\hat{H}_s,\hat{\rho}_{\chi}(t)]\\
&&-\int^{\infty}_0d{\tau}\textrm{Tr}_b\{[\hat{V}_{\chi},[\hat{V}_{\chi}(-\tau),\hat{\rho}_{\chi}(t){\otimes}\hat{\rho}_b]_{o}]_{o}\},\nonumber
\end{eqnarray}
with the commutation relation $[\hat{A}_{\chi},\hat{B}_{\chi}]_{o}=\hat{A}_{\chi}\hat{B}_{\chi}-\hat{B}_{\chi}\hat{A}_{-\chi}$
and $\hat{\rho}_{\chi}(t)$ the reduced system density operator under the counting parameter.

Then we utilize the general expression of the master equation at Eq.~(\ref{sme2}) to the present model.
Starting from the nonequilibrium spin-boson model at Eq.~({\ref{h1}}), by counting the boson number in the right bath
the Hamiltonian under the counting parameter is given by
$\hat{H}_0^{\chi}=\frac{\epsilon_0}{2}\hat{\sigma}_z+\frac{\Delta}{2}\hat{\sigma}_x+\hat{H}_B+V_0({\chi})$,
with the spin-boson interaction
\begin{equation}
\hat{V}_0({\chi})=\sigma_z\sum_{v=L,R;k}(\hat{b}_{k,v}^{\dag}e^{i\omega_k{\chi}\delta_{v,R}/2}+\hat{b}_{k,v}e^{-i\omega_k{\chi}\delta_{v,R}/2}).
\end{equation}
By applying the canonical transformation $\hat{U}_{\chi}=e^{i\hat{\sigma}_z\hat{B}_{\chi}/2}$ to $\hat{H}_0^{\chi}$,
we derive the new Hamiltonian
$\hat{H}_{\chi}=\hat{U}^{\dag}\hat{H}_0^{\chi}\hat{U}=\frac{\epsilon_0}{2}\hat{\sigma}_z+\hat{H}_B+\hat{V}(\chi)$,
where the new system-bath coupling as
$\hat{V}(\chi)=\frac{\Delta}{2}(\hat{\sigma}_x\cos\hat{B}_{\chi}+\hat{\sigma}_y\sin\hat{B}_{\chi})$,
with $\hat{B}_{\chi}=2i\sum_{k,v}(\frac{\lambda_{k,v}}{\omega_k}e^{i{\omega_k}\chi/2}\hat{b}_{k,v}^{\dag}
-\frac{\lambda^{*}_{k,v}}{\omega_k}e^{-i{\omega_k}\chi/2}\hat{b}_{k,v})$
the collective momentum of the boson baths under the counting parameter.
Since it is already known that the fluctuation of $\hat{V}(\chi)$ itself can be perturbed,
we re-group the transformed Hamiltonian as $\hat{H}_{\chi}=\hat{H}_S({\chi})+\hat{H}_B+\hat{V}^{\prime}(\chi)$,
with
\begin{equation}
\hat{H}^{\prime}_S=\frac{\epsilon_0}{2}\hat{\sigma}_z+{\eta}\frac{\Delta}{2}\hat{\sigma}_x,
\end{equation}
and
\begin{equation}
\hat{V}^{\prime}(\chi)=\frac{\Delta}{2}[\hat{\sigma}_x(\cos\hat{B}_{\chi}-\eta)+\hat{\sigma}_y\sin\hat{B}_{\chi}].
\end{equation}
The renormalization factor is shown as $\eta={\langle}\cos\hat{B}_{\chi}{\rangle}$.
For the super-Ohmic bath spectrum $J_v(\omega)={\pi}{\alpha_v}{\omega}^s{\omega_{c,v}}^{1-s}e^{-\omega/\omega_{c,v}}$ with $s=3$, it is explicitly expressed as
\begin{equation}
\eta=\exp\{\sum_{v=L,R}-\frac{\alpha_v}{2}[-1+\frac{2}{(\beta_v\omega_{c,v})^2}\psi_1(\frac{1}{\beta_v\omega_{c,v}})]\},
\end{equation}
with $\beta_v=1/k_BT_v$ the inverse temperature, $\alpha_v$ the coupling strength and $\omega_{c,v}$ the cutoff frequency of the $v$ bath, respectively. The special function $\psi_1(x)=\sum^{\infty}_{n=0}\frac{1}{(n+x)^2}$ is the trigamma function.
It should be noted that the renormalization factor is independent on the counting parameter.

Based on Eq.~(\ref{sme2}), the nonequilibrium polaron-transformed Redfield equation combined with the counting parameter is derived by
\begin{eqnarray}~\label{sme3}
\frac{{\partial}\hat{\rho}_{\chi}}{{\partial}t}&=&\frac{1}{i}[\hat{H}^{\prime}_S,\hat{\rho}_{\chi}]
-\sum_{l=e,o;\omega,\omega^{\prime}}[\Gamma_{l,+}(\omega)\hat{P}_l(\omega^{\prime})\hat{P}_l(\omega)\hat{\rho}_{\chi}\nonumber\\
&&+\Gamma_{l,-}(\omega)\hat{\rho}_{\chi}\hat{P}_l(\omega)\hat{P}_l(\omega^{\prime})
-\Gamma^{\chi}_{l,-}(\omega)\hat{P}_l(\omega^{\prime})\hat{\rho}_{\chi}\hat{P}_l(\omega)\nonumber\\
&&-\Gamma^{\chi}_{l,+}(\omega)\hat{P}_l(\omega)\hat{\rho}_{\chi}\hat{P}_l(\omega^{\prime})],
\end{eqnarray}
with $\hat{\rho}_{\chi}$ the two-level system density operator under the counting field.
$\Gamma^{\chi}_{l,\pm}(\omega)$ are the transition rates expressed by
\begin{eqnarray}~\label{cf1}
\Gamma^{\chi}_{e,\pm}(\omega)&=&(\frac{\Delta}{2})^2\int^{\infty}_0d{\tau}e^{i{\omega}\tau}
[{\langle}\cos\hat{B}_{-\chi}(\pm\tau)\cos\hat{B}_{\chi}(0){\rangle}-\eta^2]\nonumber\\
\Gamma^{\chi}_{o,\pm}(\omega)&=&(\frac{\Delta}{2})^2\int^{\infty}_0d{\tau}e^{i{\omega}\tau}
{\langle}\sin\hat{B}_{-\chi}(\pm\tau)\sin\hat{B}_{\chi}(0){\rangle},
\end{eqnarray}
and $\Gamma_{l,\pm}(\omega)=\Gamma^{\chi=0}_{l,\pm}(\omega)$.
$P_l(\omega)$ is the eigen-state projector from the evolution of the Pauli operators as
$\sigma_{x(y)}(-\tau)=\sum_{\omega=0,\pm{\Lambda}}\hat{P}_{e(o)}(\omega)e^{i{\omega}\tau}$ with the energy gap
$\Lambda=\sqrt{\epsilon_0^2+\eta^2\Delta^2}$, the eigen-states
\begin{eqnarray}
|+{\rangle}&=&\cos\frac{\theta}{2}|1{\rangle}+\sin\frac{\theta}{2}|0{\rangle},\nonumber\\
|-{\rangle}&=&-\sin\frac{\theta}{2}|1{\rangle}+\cos\frac{\theta}{2}|0{\rangle},
\end{eqnarray}
and $\tan\theta=\eta\Delta/\epsilon_0$.
In the even branch, they are specified as
$\hat{P}_e(0)=\sin\theta\hat{\tau}_z$, $\hat{P}_e(\Lambda)=\cos\theta\hat{\tau}_-$,
and $\hat{P}_e(-\Lambda)=\cos\theta\hat{\tau}_+$,
with $\hat{\tau}_z=|+{\rangle}{\langle}+|-|-{\rangle}{\langle}-|$ and $\hat{\tau}_{\pm}=|\pm{\rangle}{\langle}\mp|$.
Similarly in the odd branch, they become
$\hat{P}_o(0)=0$, $\hat{P}_o(\Lambda)=i\hat{\tau}_-$ and $\hat{P}_o(-\Lambda)=-i\hat{\tau}_+$.

A similar polaron transformation based quantum master equation has been previously proposed by R. J. Silbey \emph{et al}.~\cite{rjsilbey1,rjsilbey11} in the single bath spin-boson model under the Born-Markov approximation.
Then, it was extended to the non-Markovian regime by S. Jang \emph{et al}. to include the slow bath effect~\cite{sjangs1}.
Recently, it was confirmed both from the equilibrium statistics and quantum dynamics that the markovian master equation combined with the polaron transformation can be accurately utilized in the fast bath regime~\cite{sleeck1,sleeck2,schanght1}. We focus on the fast bath regime so that the Markovian  master equation is applied in the main text.

\section{Parity based correlation functions} \label{sectionB}
From the quantum master equation including the counting parameter at Eq.~(\ref{sme3}), it is known that the transition rates at Eq.~(\ref{cf1}) are crucial to obtain the evolution of the two-level system density matrix.
They can be straightforwardly re-expressed as
$\Gamma^{\chi}_{l,\pm}=(\frac{\eta\Delta}{2})^2\int^{\infty}_0d{\tau}e^{i{\omega}\tau}\gamma^{\chi}_{l}(\pm\tau)$,
with the real-time correlation functions being
\begin{eqnarray}
\gamma^{\chi}_{e}(\pm\tau)&=&{\cosh}Q(\pm\tau-\chi)-1=\sum^{\infty}_{n=1}\frac{Q(\pm\tau-\chi)^{2n}}{(2n)!},\nonumber\\
\gamma^{\chi}_{o}(\pm\tau)&=&{\sinh}Q(\pm\tau-\chi)=\sum^{\infty}_{n=0}\frac{Q(\pm\tau-\chi)^{2n+1}}{(2n+1)!}.
\end{eqnarray}
The boson propagator $Q(\tau-\chi)=Q_L(\tau)+Q_R(\tau-\chi)$ describes the spontaneous absorption and emission of the bosons
with
\begin{equation}
Q_v(\tau)=\int^{\infty}_0d{\omega}\frac{J_v(\omega)}{\pi\omega^2}[n_v(\omega)e^{i{\omega}\tau}+(1+n_v(\omega))e^{-i{\omega}\tau}].
\end{equation}
In absence of the counting parameter, these correlation functions reduce to Eq.~(\ref{q1}).
Then, the corresponding transition rates clearly uncover the energy transfer processes shown in Fig.~\ref{fig1}.
For the super-Ohmic bath, it can be explicitly expressed by
\begin{eqnarray}
Q_v(\tau)&=&{\alpha_v}\{[\frac{-1+\omega_{c,v}^2\tau^{2}}{(1+\omega_{c,v}^2\tau^{2})^2}
+\frac{2\textrm{Re}[\psi_1(\frac{1}{\beta_v\omega_{c,v}}+\frac{i\tau}{\beta_v})]}{(\beta_v\omega_{c,v})^2}]\nonumber\\
&&-i\frac{2\omega_{c,v}\tau}{(1+\omega_{c,v}^2\tau^{2})^2}\}.
\end{eqnarray}
Moreover, it is found that parity-cross branch ${\langle}\cos\hat{B}_{\pm\chi}\sin\hat{B}_{\chi}{\rangle}=0$.
This clearly shows that the multiple boson-assisted energy transfer is protected by the parity.
To investigate the dynamical processes of the boson-assisted energy transfer based on the correlation functions,
the frequency-domain rates are introduced by
\begin{equation}
\phi^{\chi}_{l}(\omega)=\int^{\infty}_{-\infty}d{\tau}e^{i\omega\tau}\gamma^{\chi}_l(\tau),
\end{equation}
which returns to $\phi_l(\omega)=\int^{\infty}_{-\infty}d{\tau}e^{i\omega\tau}\gamma_l(\tau)$ in the case of $\chi=0$.
It can also expanded by the kernel functions as
$\phi^{\chi}_l(\omega)=\frac{1}{2\pi}\int^{\infty}_{-\infty}d{\omega}^{\prime}e^{i\chi\omega^{\prime}}C_l(\omega,\omega^{\prime})$,
where
\begin{eqnarray}
C_e(\omega,\omega^{\prime})&=&\frac{1}{2}\sum_{\sigma={\pm}}C^{\sigma}_{L}(\omega-\omega^{\prime})C^{\sigma}_R(\omega^{\prime})-\delta(\omega^{\prime}),\nonumber\\
C_o(\omega,\omega^{\prime})&=&\frac{1}{2}\sum_{\sigma=\pm}{\sigma}C^{\sigma}_L(\omega-\omega^{\prime})C^{\sigma}_R(\omega^{\prime}),
\end{eqnarray}
with
$C^{\pm}_v(\omega^{\prime})=\int^{\infty}_{-\infty}d{\tau}e^{i{\omega^{\prime}}\tau{\pm}Q_v(\tau)}$.

\section{Definition of the steady state energy flux}\label{sectionC}
Starting from the full counting statistics, we apply the cumulant generating function to derive the energy flux into the right bath.
In the Liouville space, the quantum master equation at Eq.~(\ref{sme2}) can be re-expressed as~\cite{swangc1,syuget1}
\begin{equation}~\label{sle1}
\frac{d}{dt}|\rho_{\chi}(t){\rangle}{\rangle}=-\hat{\mathcal{L}}_{\chi}|\rho_{\chi}(t){\rangle}{\rangle},
\end{equation}
with $|\rho_{\chi}(t){\rangle}{\rangle}=[{\langle}1|\hat{\rho}_{\chi}(t)|1{\rangle},{\langle}0|\hat{\rho}_{\chi}(t)|0{\rangle},{\langle}1|\hat{\rho}_{\chi}(t)|0{\rangle},{\langle}0|\hat{\rho}_{\chi}(t)|1{\rangle}]^{T}$ in the vector form, and $\hat{\mathcal{L}}_{\chi}$ the super-operator.
The generating function is obtained by
\begin{equation}
\mathcal{Z}_{\chi}(t)=\textrm{Tr}\{\hat{\rho}_{\chi}(t)\}={\langle}{\langle}1|\hat{T}[e^{-\hat{\mathcal{L}}_{\chi}\tau}]|\rho(0){\rangle}{\rangle},
\end{equation}
where ${\langle}{\langle}1|=[1,1,0,0]$, $\hat{T}$ is the time-ordering operator and $|\rho(0){\rangle}{\rangle}$ is the initial system density matrix.
Energy transfer behaviors in the long time limit are of our prime interest in the present paper. They are controlled by the ground state of $\hat{\mathcal{L}}_{\chi}(t)$, with the ground state energy as $E_0(\chi)$ having the smallest real part.
Hence, the generating function is simplified to $\mathcal{Z}_{\chi}=\exp[-E_0(\chi)t]$.
Then the steady state cumulant generating function can be derived by
$\mathcal{G}_{\chi}=\lim_{t{\rightarrow}\infty}{\ln\mathcal{Z}_{\chi}(t)}/{t}=-E_0(\chi)$,
which finally generates the steady energy flux as $\mathcal{J}=\frac{{\partial}E_0(\chi)}{{\partial}(i\chi)}|_{\chi=0}$.

Alternatively, based on the Eq.~(\ref{sle1}) the steady state solution can be expressed as
$\hat{\mathcal{L}}_{\chi}|\Psi_{\chi}{\rangle}{\rangle}=E_0(\chi)|\Psi_{\chi}{\rangle}{\rangle}$, with $|\Psi_{\chi}{\rangle}{\rangle}$ the corresponding right ground state. Taking the derivative of $i\chi$ at two sides results in
\begin{equation}
\frac{{\partial}\hat{\mathcal{L}}_{\chi}}{{\partial}(i\chi)}|\Psi_{\chi}{\rangle}{\rangle}+\hat{\mathcal{L}}_{\chi}\frac{{\partial}|\Psi_{\chi}{\rangle}{\rangle}}
{{\partial}(i\chi)}=\frac{{\partial}E_0(\chi)}{{\partial}(i\chi)}|\Psi_{\chi}{\rangle}{\rangle}+E_0(\chi)\frac{{\partial}|\Psi_{\chi}{\rangle}{\rangle}}{{\partial}(i\chi)}.
\end{equation}
When $\chi=0$, it is known that $E_0=0$, ${\langle}{\langle}1|\hat{\mathcal{L}}=0$ and ${\langle}{\langle}1|\Psi{\rangle}{\rangle}=1$.
As a result, $\frac{{\partial}E_0(\chi)}{{\partial}(i\chi)}|_{\chi=0}={\langle}{\langle}1|\frac{{\partial}\hat{\mathcal{L}_{\chi}}}{{\partial}(i\chi)}|_{\chi=0}|\Psi{\rangle}{\rangle}$ and the energy flux is re-expressed as
\begin{equation}~\label{jdyn1}
\mathcal{J}={\langle}{\langle}1|\frac{{\partial}\hat{\mathcal{L}_{\chi}}}{{\partial}(i\chi)}|_{\chi=0}|\Psi{\rangle}{\rangle}.
\end{equation}

\section{Analytical solution of the unified energy flux}\label{sectionD}
From the expression of the energy flux at Eq.~(\ref{jdyn1}), we should first obtain the operator $\hat{\mathcal{L}}_{\chi}$.
The corresponding matrix elements from Eq.~(\ref{sle1}) are described by
\begin{widetext}
\begin{eqnarray}
\textrm{L}_{11,11}&=&\sin^2\theta[\Gamma_{e,+}(0)+\Gamma_{e,-}(0)]
+\cos\theta\cos^2\frac{\theta}{2}[\Gamma_{e,+}(\Lambda)+\Gamma_{e,-}(-\Lambda)]
-\cos\theta\sin^2\frac{\theta}{2}[\Gamma_{e,+}(-\Lambda)+\Gamma_{e,-}(\Lambda)]\nonumber\\
&&+\cos^2\frac{\theta}{2}[\Gamma_{o,+}(\Lambda)+\Gamma_{o,-}(-\Lambda)]+\sin^2\frac{\theta}{2}[\Gamma_{o,+}(-\Lambda)+\Gamma_{o,-}(\Lambda)]\nonumber
\end{eqnarray}
\begin{eqnarray}
\textrm{L}_{11,00}&=&-\sin^2\theta[\Gamma^{\chi}_{e,+}(0)+\Gamma^{\chi}_{e,-}(0)]
-\cos\theta\cos^2\frac{\theta}{2}[\Gamma^{\chi}_{e,+}(-\Lambda)+\Gamma^{\chi}_{e,-}(\Lambda)]
+\cos\theta\sin^2\frac{\theta}{2}[\Gamma^{\chi}_{e,+}(\Lambda)+\Gamma^{\chi}_{e,-}(-\Lambda)]\nonumber\\
&&-\cos^2\frac{\theta}{2}[\Gamma^{\chi}_{o,+}(-\Lambda)+\Gamma^{\chi}_{o,-}(\Lambda)]
-\sin^2\frac{\theta}{2}[\Gamma^{\chi}_{o,+}(\Lambda)+\Gamma^{\chi}_{o,-}(-\Lambda)]\nonumber
\end{eqnarray}
\begin{eqnarray}
\textrm{L}_{11,10}&=&\frac{\sin\theta}{2}([\Gamma^{\chi}_{o,+}(-\Lambda)+\Gamma_{o,-}(-\Lambda)]-[\Gamma^{\chi}_{o,+}(\Lambda)+\Gamma_{o,-}(\Lambda)])\nonumber\\
&&+\frac{\sin{2\theta}}{2}(\frac{1}{2}[\Gamma^{\chi}_{e,+}(-\Lambda)+\Gamma_{e,-}(\Lambda)+\Gamma^{\chi}_{e,+}(\Lambda)
+\Gamma_{e,-}(-\Lambda)]-[\Gamma^{\chi}_{e,+}(0)+\Gamma_{e,-}(0)])\nonumber
\end{eqnarray}
\begin{eqnarray}
\textrm{L}_{11,01}&=&\frac{\sin\theta}{2}([\Gamma_{o,+}(\Lambda)+\Gamma^{\chi}_{o,-}(\Lambda)]-[\Gamma_{o,+}(-\Lambda)+\Gamma^{\chi}_{o,-}(-\Lambda)])\nonumber\\
&&+\frac{\sin{2\theta}}{2}(\frac{1}{2}[\Gamma_{e,+}(-\Lambda)+\Gamma^{\chi}_{e,-}(\Lambda)
+\Gamma_{e,+}(\Lambda)+\Gamma^{\chi}_{e,-}(-\Lambda)]-[\Gamma_{e,+}(0)+\Gamma^{\chi}_{e,-}(0)])\nonumber
\end{eqnarray}
\begin{eqnarray}
\textrm{L}_{00,11}&=&-\sin^2{\theta}[\Gamma^{\chi}_{e,+}(0)+\Gamma^{\chi}_{e,-}(0)]
-\cos\theta\cos^2\frac{\theta}{2}[\Gamma^{\chi}_{e,+}(\Lambda)+\Gamma^{\chi}_{e,-}(-\Lambda)]
+\cos\theta\sin^2\frac{\theta}{2}[\Gamma^{\chi}_{e,+}(-\Lambda)
+\Gamma^{\chi}_{e,-}(\Lambda)]\nonumber\\
&&-\cos^2\frac{\theta}{2}[\Gamma^{\chi}_{o,+}(\Lambda)+\Gamma^{\chi}_{o,-}(-\Lambda)]
-\sin^2\frac{\theta}{2}[\Gamma^{\chi}_{o,+}(-\Lambda)+\Gamma^{\chi}_{o,-}(\Lambda)]\nonumber
\end{eqnarray}
\begin{eqnarray}
\textrm{L}_{00,00}&=&\sin^2\theta[\Gamma_{e,+}(0)+\Gamma_{e,-}(0)]
+\cos\theta\cos^2\frac{\theta}{2}[\Gamma_{e,+}(-\Lambda)+\Gamma_{e,-}(\Lambda)]
-\cos\theta\sin^2\frac{\theta}{2}[\Gamma_{e,+}(\Lambda)+\Gamma_{e,-}(-\Lambda)]\nonumber\\
&&+\cos^2\frac{\theta}{2}[\Gamma_{o,+}(-\Lambda)+\Gamma_{o,-}(\Lambda)]
+\sin^2\frac{\theta}{2}[\Gamma_{o,+}(\Lambda)+\Gamma_{o,-}(-\Lambda)]\nonumber
\end{eqnarray}
\begin{eqnarray}
\textrm{L}_{00,10}&=&\frac{\sin\theta}{2}([\Gamma_{o,+}(\Lambda)+\Gamma^{\chi}_{o,-}(\Lambda)]
-[\Gamma_{o,+}(-\Lambda)+\Gamma^{\chi}_{o,-}(-\Lambda)])\nonumber\\
&&+\frac{\sin2\theta}{2}([\Gamma_{e,+}(0)+\Gamma^{\chi}_{e,-}(0)]-\frac{1}{2}[\Gamma^{\chi}_{e,-}(\Lambda)+\Gamma_{e,+}(-\Lambda)+\Gamma^{\chi}_{e,-}(-\Lambda)+\Gamma_{e,+}(\Lambda)])\nonumber\\
\end{eqnarray}
\begin{eqnarray}
\textrm{L}_{00,01}&=&\frac{\sin\theta}{2}([\Gamma_{o,-}(-\Lambda)+\Gamma^{\chi}_{o,+}(-\Lambda)]
-[\Gamma_{o,-}(\Lambda)+\Gamma^{\chi}_{o,+}(\Lambda)])\nonumber\\
&&+\frac{\sin{2\theta}}{2}([\Gamma_{e,-}(0)+\Gamma^{\chi}_{e,+}(0)]
-\frac{1}{2}[\Gamma^{\chi}_{e,+}(\Lambda)+\Gamma_{e,-}(-\Lambda)
+\Gamma^{\chi}_{e,+}(-\Lambda)+\Gamma_{e,-}(\Lambda)])\nonumber
\end{eqnarray}
\begin{eqnarray}
\textrm{L}_{10,11}&=&\frac{\sin\theta}{2}([\Gamma^{\chi}_{o,+}(\Lambda)-\Gamma_{o,-}(\Lambda)]+[\Gamma_{o,-}(-\Lambda)-\Gamma^{\chi}_{o,+}(-\Lambda)])\nonumber\\
&&+\frac{\sin2\theta}{2}([\Gamma_{e,-}(0)-\Gamma^{\chi}_{e,+}(0)]+\frac{1}{2}[\Gamma^{\chi}_{e,+}(\Lambda)-\Gamma_{e,-}(\Lambda)
+\Gamma^{\chi}_{e,+}(-\Lambda)-\Gamma_{e,-}(-\Lambda)])\nonumber
\end{eqnarray}
\begin{eqnarray}
\textrm{L}_{10,00}&=&\frac{\sin\theta}{2}([\Gamma_{o,+}(\Lambda)-\Gamma^{\chi}_{o,-}(\Lambda)]+[\Gamma^{\chi}_{o,-}(-\Lambda)-\Gamma_{o,+}(-\Lambda)])\nonumber\\
&&+\frac{\sin2\theta}{2}([\Gamma^{\chi}_{e,-}(0)-\Gamma-{e,+}(0)]
+\frac{1}{2}[\Gamma_{e,+}(\Lambda)-\Gamma^{\chi}_{e,-}(\Lambda)+\Gamma_{e,+}(-\Lambda)-\Gamma^{\chi}_{e,-}(-\Lambda)])\nonumber
\end{eqnarray}
\begin{eqnarray}
\textrm{L}_{10,10}&=&\sin^2\theta[\Gamma_{e,+}(0)+\Gamma_{e,-}(0)]+\cos\theta\cos^2\frac{\theta}{2}[\Gamma_{e,+}(\Lambda)
+\Gamma_{e,-}(\Lambda)]-\cos\theta\sin^2\frac{\theta}{2}[\Gamma_{e,+}(-\Lambda)+\Gamma_{e,-}(-\Lambda)]\nonumber\\
&&+\cos^2\frac{\theta}{2}[\Gamma_{o,+}(\Lambda)+\Gamma_{o,-}(\Lambda)]
+\sin^2\frac{\theta}{2}[\Gamma_{o,+}(-\Lambda)+\Gamma_{o,-}(-\Lambda)]\nonumber
\end{eqnarray}
\begin{eqnarray}
\textrm{L}_{10,01}&=&-\sin^2\theta[\Gamma^{\chi}_{e,-}(0)+\Gamma^{\chi}_{e,+}(0)]+\cos\theta\sin^2\frac{\theta}{2}
[\Gamma^{\chi}_{e,-}(\Lambda)+\Gamma^{\chi}_{e,+}(\Lambda)]-\cos\theta\cos^2\frac{\theta}{2}[\Gamma^{\chi}_{e,-}(-\Lambda)
+\Gamma^{\chi}_{e,+}(-\Lambda)]\nonumber\\
&&+\sin^2\frac{\theta}{2}[\Gamma^{\chi}_{o,+}(\Lambda)+\Gamma^{\chi}_{o,-}(\Lambda)]
+\cos^2\frac{\theta}{2}[\Gamma^{\chi}_{o,+}(-\Lambda)+\Gamma^{\chi}_{o,-}(-\Lambda)]\nonumber
\end{eqnarray}
\begin{eqnarray}
\textrm{L}_{01,11}&=&\frac{\sin\theta}{2}([\Gamma_{o,+}(\Lambda)-\Gamma^{\chi}_{o,-}(\Lambda)]
+[\Gamma^{\chi}_{o,-}(-\Lambda)-\Gamma_{o,+}(-\Lambda)])\nonumber\\
&&+\frac{{\sin}2\theta}{2}([\Gamma_{e,+}(0)-\Gamma^{\chi}_{e,-}(0)]
+\frac{1}{2}[\Gamma^{\chi}_{e,-}(\Lambda)-\Gamma_{e,+}(\Lambda)+\Gamma^{\chi}_{e,-}(-\Lambda)-\Gamma_{e,+}(-\Lambda)])\nonumber
\end{eqnarray}
\begin{eqnarray}
\textrm{L}_{01,00}&=&\frac{\sin\theta}{2}([\Gamma^{\chi}_{o,+}(\Lambda)-\Gamma_{o,-}(\Lambda)]+[\Gamma_{o,-}(-\Lambda)-\Gamma^{\chi}_{o,+}(-\Lambda)])\nonumber\\
&&+\frac{{\sin}2\theta}{2}([\Gamma^{\chi}_{e,+}(0)-\Gamma_{e,-}(0)]
+\frac{1}{2}[\Gamma_{e,-}(\Lambda)-\Gamma^{\chi}_{e,+}(\Lambda)+\Gamma_{e,-}(-\Lambda)-\Gamma^{\chi}_{e,+}(-\Lambda)])\nonumber
\end{eqnarray}
\begin{eqnarray}
\textrm{L}_{01,10}&=&-\sin^2\theta[\Gamma^{\chi}_{e,+}(0)+\Gamma^{\chi}_{e,-}(0)]-\cos\theta\cos^2\frac{\theta}{2}
[\Gamma^{\chi}_{e,+}(\Lambda)+\Gamma^{\chi}_{e,-}(\Lambda)]+\cos\theta\sin^2\frac{\theta}{2}[\Gamma^{\chi}_{e,+}(-\Lambda)+\Gamma^{\chi}_{e,-}(-\Lambda)]\nonumber\\
&&+\cos^2\frac{\theta}{2}[\Gamma^{\chi}_{o,+}(\Lambda)+\Gamma^{\chi}_{o,-}(\Lambda)]
+\sin^2\frac{\theta}{2}[\Gamma^{\chi}_{o,+}(-\Lambda)+\Gamma^{\chi}_{o,-}(-\Lambda)]\nonumber
\end{eqnarray}
\begin{eqnarray}
\textrm{L}_{01,01}&=&\sin^2\theta[\Gamma_{e,+}(0)+\Gamma_{e,-}(0)]+\cos\theta\cos^2\frac{\theta}{2}[\Gamma_{e,+}(-\Lambda)
+\Gamma_{e,-}(-\Lambda)]-\cos\theta\sin^2\frac{\theta}{2}[\Gamma_{e,+}(\Lambda)+\Gamma_{e,-}(\Lambda)]\nonumber\\
&&+\cos^2\frac{\theta}{2}[\Gamma_{o,+}(-\Lambda)+\Gamma_{o,-}(-\Lambda)]
+\sin^2\frac{\theta}{2}[\Gamma_{o,+}(\Lambda)+\Gamma_{o,-}(\Lambda)],\nonumber
\end{eqnarray}
\end{widetext}
where $\textrm{L}_{ij,kl}={\langle}{\langle}ij|\hat{\mathcal{L}}_{\chi}|kl{\rangle}{\rangle}$ with the basis
$|ij{\rangle}{\rangle}=|i{\rangle}{\langle}j|$ in the Hilbert space, and the transition rates $\Gamma^{\chi}_{l,\pm}(\omega)$ are given at Eq.~(\ref{cf1}).
Combined with $|\Psi{\rangle}{\rangle}=[P_{11},P_{00},P_{10},P_{01}]^T$ with $P_{ij}={\langle}i|\hat{\rho}(\infty)|j{\rangle}$ the density matrix element at steady state, the energy flux can be expressed as
\begin{eqnarray}
\mathcal{J}&=&-[\frac{{\partial}\textrm{L}_{00,11}(\chi)}{{\partial}(i\chi)}|_{\chi=0}P_{11}+\frac{{\partial}\textrm{L}_{11,00}(\chi)}{{\partial}(i\chi)}|_{\chi}P_{00}\nonumber\\
&&+\frac{{\partial}\textrm{L}_{10}(\chi)}{{\partial}(i\chi)}|_{\chi=0}P_{10}
+\frac{{\partial}\textrm{L}_{01}(\chi)}{{\partial}(i\chi)}|_{\chi=0}P_{01}],
\end{eqnarray}
with $\textrm{L}_{10(01)}(\chi)=\textrm{L}_{11,10(01)}(\chi)+\textrm{L}_{00,10(01)}(\chi)$.
The case of $\epsilon_0=0$ is of the prime interest, corresponding to $\theta=\pi/2$.
Then the crucial transition rates are simplified to
\begin{eqnarray}
\textrm{L}_{00,11}(\chi)&=&-\frac{\Lambda^{2}}{4}(\phi^{\chi}_e(0)+\frac{1}{2}[\phi^{\chi}_o(\Lambda)+\phi^{\chi}_o(-\Lambda)]), \nonumber\\
\textrm{L}_{10}(\chi)&=&\frac{\Lambda^{2}}{8}([\phi_o(\Lambda)-\phi_o(-\Lambda)]-[\phi^{\chi}_o(\Lambda)-\phi^{\chi}_o(-\Lambda)]),\nonumber
\end{eqnarray}
with the two-level system energy gap $\Lambda=\eta\Delta$, $\textrm{L}_{11,00}(\chi)=\textrm{L}_{00,11}(\chi)$, and
$\textrm{L}_{01}(\chi)=\textrm{L}_{10}(\chi)$.
Moreover, the steady state solution of the Eq.~(\ref{sle1}) in absence of the counting parameter ($\chi=0$) is given by
\begin{equation}
P_{11}=P_{00}=1/2,~P_{10}=P_{01}=\frac{1}{2}\frac{\phi_o(-\Lambda)-\phi_o(\Lambda)}{\phi_o(-\Lambda)+\phi_o(\Lambda)}.
\end{equation}
Finally, the energy flux is obtained by
\begin{eqnarray}
\mathcal{J}&=&\frac{{\Lambda}^2}{8\pi}\int^{\infty}_{-\infty}d{\omega}
\bigg[\frac{\phi_o(\Lambda)C_o(-\Lambda,\omega)+\phi_o(-\Lambda)C_o(\Lambda,\omega)}{\phi_o(\Lambda)+\phi_o(-\Lambda)}\nonumber\\
&&+C_e(0,\omega)\bigg]{\omega}.\nonumber\\
\end{eqnarray}



\end{document}